\documentclass[rnote]{aa}
\usepackage{graphicx}
\usepackage{txfonts}

\begin{document}

\title{Studying the properties of the radio emitter in LS~5039}

\author{V. Bosch-Ramon\inst{1}} 

\institute{Max Planck Institut f\"ur Kernphysik, Saupfercheckweg 1, Heidelberg 69117, Germany; 
vbosch@mpi-hd.mpg.de}

\offprints{ \\ \email{vbosch@mpi-hd.mpg.de}}

\abstract
{LS~5039 is an X-ray binary that presents non-thermal radio emission. The radiation at $\sim 5$~GHz is quite steady and
optically thin, consisting on a dominant core plus an extended jet-like structure. There is a spectral turnover around 1~GHz,
and evidence of variability at timescales of 1~yr at 234~MHz.}
{We investigate the radio emitter properties using the available broadband radio data, and assuming two possible scenarios to
explain the turnover: free-free absorption in the stellar wind, or synchrotron self-absorption.}
{We use the relationships between the turnover frequency, the stellar wind density, the emitter location, size and magnetic
field, and the Lorentz factor of the emitting electrons, as well as a reasonable assumption on the energy budget, to infer
the properties of the low-frequency radio emitter. Also, we put this information in context with the broadband radio data.}
{The location and size of the low-frequency radio emitter can be restricted to $\ga$~few~AU from the primary star, its
magnetic field to $\sim 3\times 10^{-3}-1$~G,  and the electron Lorentz factors to $\sim 10-100$.  The observed variability
of the extended structures seen with VLBA would point to electron bulk velocities $\ga 3\times 10^8$~cm~s$^{-1}$, whereas 
much less variable radiation at 5~GHz
would indicate velocities for the VLBA core $\la 10^8$~cm~s$^{-1}$. The emission at
234~MHz in the high state would mostly come from a region larger than the dominant broadband radio emitter.}
{We suggest a scenario in which secondary pairs, created via gamma-ray absorption and moving in the stellar wind, are behind 
the steady broadband radio core, whereas the resolved jet-like radio emission would come from a collimated, faster, outflow.}
\keywords{Radio continuum: stars -- 
X-rays: binaries -- stars: individual: LS~5039 -- Radiation mechanisms: non-thermal}

\maketitle

\section{Introduction}\label{intro}

LS~5039 is an X-ray binary that presents broadband non-thermal emission: in radio (e.g. Mart\'i et al. \cite{marti98};  M98
hereafter); in X-rays (e.g.  Bosch-Ramon et al. \cite{bosch07}); at GeV energies (Paredes et al. \cite{paredes00}; PA0); and
at TeV energies (Aharonian \cite{aharonian05,aharonian06}), as well as extended $\sim 10-1000$~AU radio jets (e.g. PA0;
Paredes et al. \cite{paredes02}; Rib\'o et al. \cite{ribo08} -R08 hereafter-). The mass of the compact object is not well
constrained (Casares et al. \cite{casares05}; C05 hereafter), and it may harbor an accreting black-hole (e.g. Paredes,
Bosch-Ramon \& Romero  \cite{paredes06}) or a non-accreting pulsar (e.g. Martocchia, Motch \& Negueruela \cite{martocchia05};
Dubus \cite{dubus06}). The source is at a distance of 2.5~kpc (C05).

Low-frequency ($\la 1$~GHz) radio emission has been detected by the Giant Metrewave Radio Telescope -GMRT- (Pandey et al.
\cite{pandey07}; Godambe et al. \cite{godambe08}; P07 and G08 hereafter) in LS~5039. This radio emission is very likely of
synchrotron origin, matching well with the non-thermal radio spectrum at higher frequencies (see fig.~2 in G08). G08 found a
spectral turnover at $\nu_{\rm max}=964\pm 104$~MHz (with a flux $F_{\rm max}\approx 40$~mJy), fluxes of $\approx 17.0\pm
1.1$ and  $34.2\pm 2.8$~mJy at 234 and 614~MHz, respectively, and a spectral index $\alpha=0.75\pm 0.11$ ($F_{\nu}\propto
\nu^{\alpha}$) in this frequency range. These data were taken simultaneously (MJD=53788.78). Above the spectral
turnover, at 1.28~GHz, the flux was $38.1\pm 2.3$~mJy, being this data point obtained 18~days later (MJD=53806.73)
than those at lower frequencies. Above the spectral turnover, $\alpha$ was $\approx -0.43$, very similar to that found by 
M98. The P07 observations, taken in two different epochs (MJD=53220 and 53377) more than 1~yr apart from the G08
observations, found significantly different fluxes at 234~MHz: $65.04\pm 7.71$ and $74.56\pm 5.63$~mJy, which is evidence of
variability. Otherwise, the fluxes of P07 at 614~MHz were $28.55\pm 3.56$ and $34.29\pm 2.34$~mJy (MJD=53220, 53377), similar
to the value of G08. The $\alpha$ found by P07 were $\approx -0.81$ and $-0.86$.

The spectrum found at low frequencies by G08, far more inverted than the one corresponding to a monoenergetic distribution of
particles (i.e. $\alpha=1/3$), could not be explained by an electron population with a high minimum energy, being the most
likely origin of the turnover either free-free absorption in the stellar wind or synchrotron self-absorption. Based on
this, we investigate in this work the properties of the low-frequency radio emitter in LS~5039 adopting a plain framework in
which the radio emission, at different narrow frequency bands, is produced in an homogeneous spherical emitter with magnetic
field $B$, size $r$, and at a distance $R\ga r$ from the star/binary system. We also derive physical constraints of the 
broadband radio emitter from data at higher frequencies. Finally, we suggest a scenario for the radio emission, and a
more complex model is applied to test its feasibility.

\section{Studying the low-frequency radio emission from LS~5039}\label{abs}

In this section, we infer basic properties of the low-frequency radio emitter in LS~5039 accounting for the  detection
of a spectral turnover at $\approx 1$~GHz and the observational evidence of variability at 234~MHz. 

\begin{figure*}[]
\begin{center}
\includegraphics[width=0.6\textwidth,angle=270]{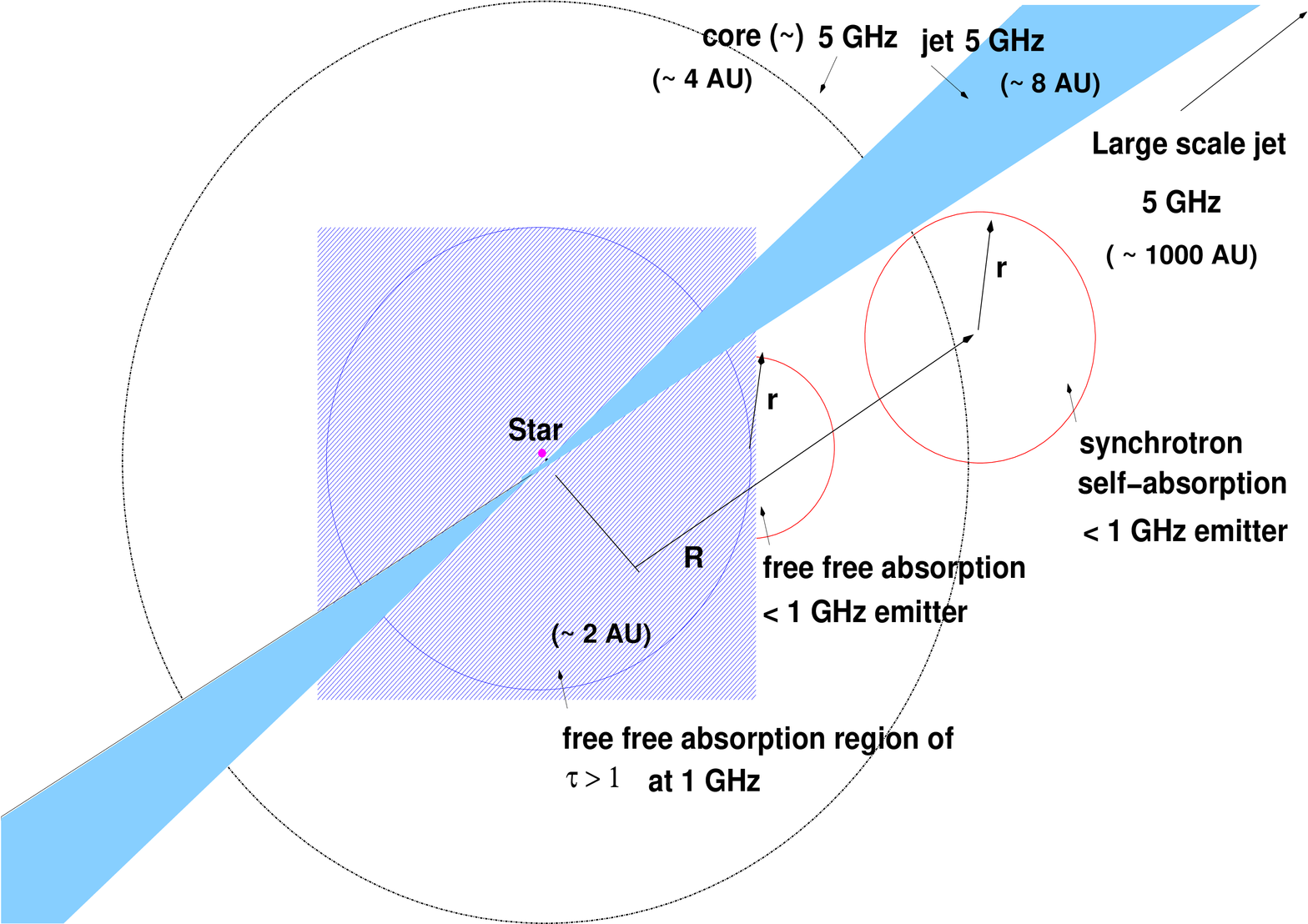}
\caption{Sketch of the regions producing radio emission in LS~5039. The main components of the radio emitter at low and 
high frequencies are shown, together with the relevant spatial scales.}
\label{plots}
\end{center}
\end{figure*}

For illustrative purposes, we present in Figure~\ref{plots} a sketch of the scenario considered here. In the picture, the
low-frequency radio emitter location is illustrated for two cases: a low-frequency turnover produced by free-free
absorption, with the emission coming from a region at few AU from the star, or by synchrotron self-absorption,
with the emitter located at similar or larger distances (see below). The wind region optically thick by free-free absorption
to 1~GHz emission is shown. The core and the jet-like emission at 5~GHz, discussed later, are also sketched.

\subsection{Free-free absorption}\label{ff}

The strongly inverted low-frequency radio spectrum could be produced by free-free absorption in the stellar wind.
The distance at which the stellar wind becomes optically thin to radio emission can be computed fixing the 
free-free opacity $\tau_{\nu~ff}$ 
(Rybicki \& Lightman \cite{rybicki79}) to 1:
\begin{equation}
\tau_{\nu~ff}\approx 30\dot{M}_{-7}^2\nu_{\rm GHz~max}^{-2}R_{\rm AU}^{-3}V_{\rm w~8.3}^{-2}T_{\rm w~4}^{-3/2}=1\,,
\end{equation}
which, for $\nu_{\rm GHz~max}=\nu_{\rm max}/{\rm 1~GHz}=1$, yields:
\begin{equation}
R\approx 3\dot{M}_{-7}^{2/3}V_{\rm w~8.3}^{-2/3}T_{\rm w~4}^{-1/2}\,{\rm AU}\,,
\end{equation}
where $\dot{M}_{-7}=\dot{M}/10^{-7}~$M$_{\odot}$~yr$^{-1}$ is the stellar mass loss rate, $R_{\rm AU}=R/1~{\rm AU}$, $V_{\rm
w~8.3}=V_{\rm w}/2\times 10^8$~cm~s$^{-1}$ the stellar wind velocity, and $T_{\rm w~4}=T_{\rm w}/10^4$~K the wind
temperature. Adopting $T_{\rm w~4}=1$ (Krtidka \& Kub\'at \cite{krtidka01}), $V_{\rm 8.3}=1$ (C05), and
$\dot{M}_{-7}=1$, we obtain $R\approx 3$~AU ($4.5\times 10^{13}$~cm; recall that $r\la R$) as the distance between the
low-frequency radio emitter and the star. 

A $\dot{M}$ somewhat lower than the values given by C05 has been taken to 
account for possible wind clumpiness in LS~5039 (Bosch-Ramon et al. \cite{bosch07}). 
Actually, wind clumpiness may prevent the wind to be an effective free-free 
absorber for large enough porosity lengths (e.g. Owocki \& Cohen \cite{owocki06}). 
In such a case, if a pulsar would be present in the system, the radio pulse may not be smeared out, hence being detectable. 
No radio pulsations have been found up to now.

If the spectral turnover is produced by wind free-free absorption, 
the synchrotron self-absorption frequency must be $<1$~GHz. Thus,
to find a restriction for $B$, one can use the following relationship between $\nu_{\rm max}$, $F_{\rm max}$, 
$r$ and $B$ in the synchrotron 
self-absorption case:
\begin{equation}
\nu_{\rm max}\approx 3.9\times 10^{10} F_{\rm max~mJy}^{2/5} B_{\rm G}^{1/5}r_{\rm cm}^{-4/5}\,{\rm GHz}\,.
\label{abssync}
\end{equation}
where $F_{\rm max~mJy}=F_{\rm max}/{\rm mJy}$, and $r_{\rm cm}=r/{\rm cm}$.
Since $F_{\rm max~mJy}=F_{\rm max}/{\rm mJy}=40$ and $r\la R\approx 4.5\times 10^{13}$~cm, 
the condition $\nu_{\rm max}<1\,{\rm GHz}$ imposes that 
$B\la 3\times 10^{-2}$~G. 

The constraint for $B$, together with 
the following formula (e.g. Pacholczyk \cite{Pacholczyk70}): 
\begin{equation}
\nu\approx 4.2\times 10^{-3}B_{\rm G}\gamma^2\,{\rm GHz}\,, 
\label{rel}
\end{equation}
allows us to constrain the Lorentz factor of the electrons emitting at 1~GHz to $\gamma\ga 10^2$. 

Further restrictions on $r$, $B$ and $\gamma$ can be derived from energy budget constraints. On one hand, the 
synchrotron cooling timescale for 1~GHz emitting electrons is:
\begin{equation}
t_{\rm sync}\approx 5\times 10^8B_{\rm G}^{-2}\gamma^{-1}\approx 3.3\times 10^7B_{\rm G}^{-3/2}\,{\rm s}\,.
\label{times}
\end{equation}
In general, radio electrons will be transported/advected out of the emitter 
well before radiating all their energy via synchrotron emission, i.e. $t_{\rm adv}\ll t_{\rm sync}$. For reasonable 
advection velocities $v_{\rm adv}\ga 2\times 10^8$~cm~s$^{-1}$ 
and emitter sizes $r\la R\approx 4.5\times 10^{13}$~cm 
(free-free absorption case), one gets $t_{\rm adv}\la 2\times 10^5$~s,
where $v_{\rm adv}$ has been taken assuming that the emitter flow cannot move slower than the stellar wind.
On the other hand,
the total broadband non-thermal luminosity of the source, $L_{\rm e}\sim 10^{36}$~erg~s$^{-1}$, can be taken as an
upper-limit for the injected luminosity in the form of radio emitting electrons. Finally, the observed low-frequency
luminosity is $L_{\rm GHz}\sim 10^{30}$~erg~s$^{-1}$. From all this, we can write:
\begin{equation}
t_{\rm sync}\la t_{\rm adv}\,(L_{\rm e}/L_{\rm GHz})\la 2\times 10^{11}\,{\rm s}\,. 
\label{ts}
\end{equation}
Therefore, accounting for Eqs.~(\ref{rel}), (\ref{times}) and (\ref{ts}), we 
derive $B\ga 3\times 10^{-3}$~G and $\gamma\la 3\times 10^2$, which allow us to 
restrict $B$ to $\sim (3-30)\times 10^{-3}$~G, and $\gamma$ to $\sim (1-3)\times 10^2$. The minimum 
$B$-value, and $\nu_{\rm max}<1\,{\rm GHz}$, imply that $r\sim (3-4.5)\times 10^{13}$~cm.

\subsection{Synchrotron self-absorption}\label{ssa}

If synchrotron self-absorption is the origin of the turnover, we can obtain, from Eqs.~(\ref{abssync}) and (\ref{rel}),
$\nu_{\rm max}=1$~GHz, and $F_{\rm max}\approx 40$~mJy, the following $r-B$ and $r-\gamma$ relationships:
\begin{equation}
r\approx 10^{14}B_{\rm G}^{1/4}\approx 4\times 10^{14}\gamma^{-1/2}\,{\rm cm}\,.
\label{rad1}
\end{equation}
Since the synchrotron emitting electrons must be relativistic, $\gamma\ga 10$, and therefore, 
$r\la 10^{14}$~cm and $B\la 1$~G. 

In this case, having a constraint on $r$, the argument of the limited energy budget can be applied as well, allowing 
us to
derive further restrictions on the parameters: $B\sim 10^{-2}-1$~G; $\gamma\sim 10-10^2$; and $r\sim (4.5-10)\times
10^{13}\,{\rm cm}\sim$~few~AU (recall $R\ga r$). 

\subsection{Variable emission at 234~MHz}\label{var}

The combination of P07 and G08 observations show variability at 234~MHz of a 400\% at year timescales, too large to be
attributed only to interstellar medium scintillation effects. The flux at 614~MHz does not show significant
changes. The value of $t_{\rm sync}$ at 234~MHz is: 
\begin{equation}
t_{\rm sync}=1.4\times 10^8B_{\rm G}^{-3/2}\,{\rm s}\,,
\end{equation}
probably too long to explain the variability just
with changes of the electron injection. Some level of variability around 1~GHz
may be induced by changes in the stellar wind if free-free absorption 
produced the spectral turnover, but at the estimated distance for the free-free absorption 
case, synchrotron self-absorption should still affect the radiation at 234~MHz due to the region compactness.

The optically thin nature of P07 data, and the fast variability, can be otherwise explained by advection with timescales
shorter than $\sim 1$~yr together with injection variations. Taking $v_{\rm adv}\la 10^{10}$~cm~s$^{-1}$, 
$\sim 1$~yr timescale restricts the size where the
(high-state) 234~MHz radiation is produced to $r\la 3\times 10^{17}$~cm. 
$v_{\rm adv}$ has been taken here to be at most mildly relativistic to avoid the complexities of significant 
boosting effects (otherwise not expected; see Paredes et al. \cite{paredes00,paredes02}).
In addition, the fact that this emission is not
synchrotron self-absorbed in the P07 data, and a flux at 234~MHz of $\approx 70$~mJy, implies a size:
\begin{equation}
r\ga 9\times 10^{14}B_{\rm G}^{1/4}\approx 4\times 10^{15}\gamma^{-1/2}\,{\rm cm}\,. 
\label{rvar}
\end{equation}

Accounting for energy budget constraints, the 1~yr variability, and the constraint $\gamma\ga 10$ (i.e. electrons must
be relativistic), to Eq.~(\ref{rvar}), we get: $B\sim 4\times 10^{-4}-0.6$~G; $\gamma\sim 10-400$; and $r\sim R\sim
10^{14}-3\times 10^{17}$~cm. $R$ has similar limits to those of $r$; if bigger, the source would have appeared as extended
under the GMRT resolution of $\sim 10$~arc-seconds at 234~MHz.

Concluding, the high-state 234~GHz emitting region probably has a larger size and smaller magnetic field than the region
producing the emission $\ge 614$~MHz and at 234~MHz in the low-state, which indicates that they are probably
different. The different spectral shape of P07 data and data $> 614$~MHz would also point to a different electron
population. Otherwise, despite the variability at 234~MHz, the quite steady $614$~MHz flux makes the conclusions of
Sect.~\ref{ff} still valid concerning the break at $\approx 1$~GHz.

In Table~\ref{tab1}, at the top, the results of all this section are summarized.

 \begin{table*}[]
  \begin{center}
  \caption[]{Basic physical parameters of the radio emitter in LS~5039}
  \label{tab1}
  \begin{tabular}{cllll}
  \hline\noalign{\smallskip}
  \hline\noalign{\smallskip}
case & $B$ & $\gamma$ & $r$ & $R$ \\
 & [G] &  & [cm] & [cm] \\
  \hline\noalign{\smallskip}
free-free abs. & $\sim (3-30)\times 10^{-3}$ & $\sim (1-8)\times 10^2$ & $\sim (3-4.5)\times 10^{13}$ & $\sim 4.5\times 10^{13}$ \\
synchrotron self-abs. & $\sim 10^{-2}-1$ & $\sim 10-10^2$ & $\sim (4.5-10)\times 10^{13}$ & $\ga (4.5-10)\times 10^{13}$ \\
234~MHz variability & $\sim 4\times 10^{-4}-0.6$ & $\sim 10-400$ & $\sim 10^{14}-3\times 10^{17}$ & 
$\sim 10^{14}-3\times 10^{17}$ \\ 
 \hline\noalign{\smallskip}
\end{tabular}  
\begin{tabular}{clll}
  \hline\noalign{\smallskip}  
Radio structure & projected radius & velocity & variability level \\ 
 & [AU] & [cm~s$^{-1}$] & \\
  \hline\noalign{\smallskip}   
radio core & $\sim 4$ & $\la 10^8$ & $\approx 10$\% \\ 
extended components & $\sim 8$ & $\ga 3\times 10^8$ & $\approx 44$\% (SE) / 260\% (NW) \\  
  \hline\noalign{\smallskip}
  \end{tabular}
  \end{center}
\end{table*}

\section{The broadband radio emission}

\subsection{The 5~GHz VLBA radio emission}

The core seen with VLBA at 5~GHz in LS~5039 is not point like, presenting a projected size of $\sim 8$~AU, and two additional
south-east (SE) and north-west (NW) components at a projected distance of $\sim 8$~AU from the core center are also detected
(PA0; R08). The variation of the core flux at 5~GHz within 5~days of difference is about a 10\%, whereas the changes in flux
of the SE and NW components are of about a 44 and a 260\%, respectively (R08). Assuming that the flux changes of the SE and
NW component fluxes are showing power engine variations (unnecessarily affecting both components in the same way), plus the
fact that the synchrotron cooling timescales of radio electrons are too long to explain such a variability, we can derive
that $v_{\rm adv}\la 10^8$~cm~s$^{-1}$ to have a steady core, and $\ga 3\times 10^8$~cm~s$^{-1}$ for the variable SE and NW
components. If shorter time variability in the SE and NW components were observed, it would increase the $v_{\rm adv}$
lower limit to $3\times 10^8 (t/5~{\rm days})^{-1}$~cm~s$^{-1}$. The main properties of the VLBA radio emission are
summarized in Table~\ref{tab1}, bottom.

\begin{figure}[]
\begin{center}
\resizebox{\hsize}{!}{\includegraphics{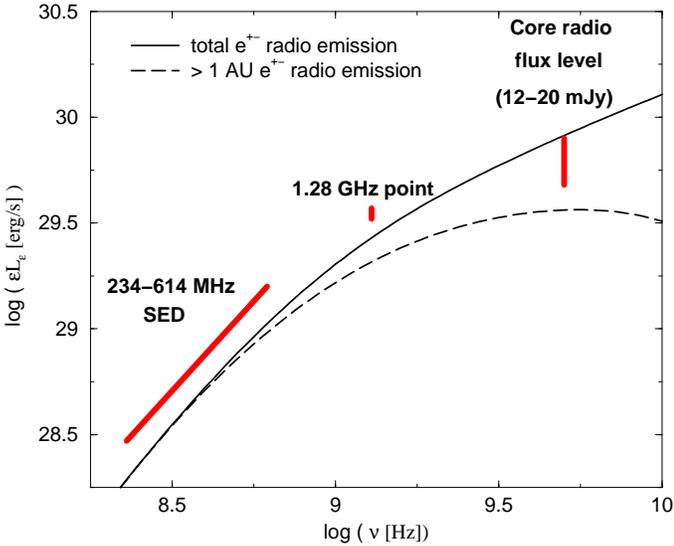}}
\caption{Computed radio SEDs of the secondary synchrotron emission created by gamma-ray absorption in LS~5039. The SEDs
corresponding  to the total emission, and to that produced at distances $\ge 1$~AU, from the star are shown. The total
differential flux at 5~GHz would be $\approx 20$~mJy and, $\ge 1$~AU, $\approx 10$~mJy. The observed luminosities from the
5~GHz core, and at 234~MHz (low state), 614~MHz and 1.28~GHz, are also shown (PA0; R08; G08).}
\label{plot2}
\end{center}
\end{figure}

Since LS~5039 harbors a very bright star (C05) and produces TeV emission (Aharonian et al. \cite{aharonian05}), secondary
pairs can be efficiently produced in the stellar wind via gamma-ray absorption. This population of secondary pairs could
be behind the radio core (e.g. Bosch-Ramon, Khangulyan \& Aharonian \cite{bosch08}; B08 hereafter), moving collectively with
the stellar wind, hence the low inferred advection speed. The feasibility of this scenario is illustrated in
Fig.~\ref{plot2}, where we show the computed radio spectral energy distribution (SED) of the secondary synchrotron emission
created by gamma-ray absorption  in LS~5039. The magnetic field in the stellar surface has been fixed to 250~G (expected to
be reasonable; see Bosch-Ramon, Khangulyan \& Aharonian \cite{bosch08b}), the injected TeV luminosity to $4\times
10^{35}$~erg~s$^{-1}$ (similar to that inferred by  Khangulyan,  Aharonian \& Bosch-Ramon \cite{khangulyan08}), and the TeV
emitter distance to the stellar companion to  $3.5\times 10^{12}$~cm, not exactly inside the binary system (as proposed
in Bosch-Ramon et al. \cite{bosch08b}). 

As seen in the Fig.~\ref{plot2}, the computed fluxes match the observed values, and most of the 5~GHz emission is
predicted to come from a region of few~AU size, similar to first order to the values discussed in the previous paragraph. The
computed  $\alpha$ around 5~GHz  is $\approx -0.4$, roughly the same as the value found by M98 for the VLA radio emission
(which should be dominated by the VLBA core). The computed low-frequency spectrum, similar to the observed one (G08),
suggests that this radiation may come also from secondary pairs, although we note that synchrotron self-absorption has been roughly 
computed
supposing an homogeneous 
emitter as wide as long. The magnetic field where most of the emission $<
1$~GHz is produced, at a distance of several AU from the star, is $\sim 1$~G, in agreement with the values  inferred in
Sect.~\ref{abs}, and also compatible with those in the wind at few AU from a massive star (Benaglia \& Romero
\cite{benaglia03}).

Concerning the SE and NW components, the larger $v_{\rm adv}$ would point to a different physical structure than that
producing the radio core, perhaps a real jet. Small changes in the orientation of $\approx 12^{\circ}$ of this jet-like
structure have been found (R08). This might be explained, for instance, by an asymmetry introduced by the secondary pairs
transported in the stellar wind (B08), the propagation of which might not be spherical, or by jet-stellar wind interactions
(Perucho \& Bosch-Ramon \cite{perucho08}).

\subsection{The jet-like structure and the 234~MHz variability}

About a 30\% of the radiation at 5~GHz comes from a jet-like structure of projected size $\sim 10-1000$~AU detected with
VLBA, MERLIN and EVN (PA0; Paredes et al. \cite{paredes02}; R08). These extended structures may be associated with the
234~MHz high-state emission, due to the size constraints for this region inferred above. 

The variability seen at 234~MHz may be related to changes in the outflow properties (e.g. via hydrodynamical
instabilities), but not to long-term variations in the primary power supply, since it should affect as well the core radio
emission. In this framework, the 234~MHz high state should have associated flux variations at higher frequencies visible
outside the radio core. Given that $t_{\rm adv}$ depends on the size and $v_{\rm adv}$, the MERLIN and EVN $\sim
100-1000$~AU structures at 5~GHz would show variability at timescales of at least months. This 234~MHz/large scale jet
link can be tested with simultaneous radio VLBI observations at low and high frequencies taken in two different epochs. We
note that a radio emitter of $r\sim 1000$~AU varying with $\sim 1$~yr timescales would have associated a $v_{\rm adv}\ga
10^9$~cm~s$^{-1}$, compatible with the $v_{\rm adv}$ derived for the VLBA SE and NW components.

\section{Discussion}

A plausible scenario for the radio emission in LS~5039 would be radiation from secondary pairs created via gamma-ray
absorption (see B08), for the compact emitter or core ($R\sim$~few~AU), plus a collimated jet, for the resolved emission. 

The jet radio emission from the smaller spatial scales would be probably flat, as typically expected in low-hard state
microquasars (e.g. Fender \cite{fender01}), but the optically thin radiation from secondary pairs would overcome this
component, at least up to few AU. This would explain why the radio spectrum of LS~5039 does not look flat (M98).  As
observed, this compact radio emission would not present significant changes along the orbit, provided first that the system
is not particularly eccentric and, second, that secondary pairs emit radio emission in a region big enough ($\sim$~few~AU) to
smooth out any possible modulation of the primary gamma-ray injection. On the other hand, the VLBA, MERLIN and EVN extended
radio emission would come from collimated and steady jets. In short, most of the $\ga 614$~MHz and the low-state 234~MHz
emission would come from the secondary pair radiation, forming the radio core, whereas the high-state 234~MHz emission, and
the 30\% of the whole 5~GHz radiation, would come from the extended jets. 

The 10--1000~AU collimated jets would be consistent with a microquasar nature for LS~5039, and hard to reconcile with
the pulsar cometary-tail model (Dubus \cite{dubus06}; Dhawan et al. \cite{dhawan06}), although further theoretical work is
required before discarding out the pulsar scenario, given the complexities inherent to hydrodynamical and
magnetohydrodynamical flows (for recent simulations on pulsar winds, see, e.g., Romero et al. \cite{romero07}; Bogovalov et
al. \cite{bogovalov08}). 

The scenario proposed here can be tested looking for: variability in the extended emission at scales $\sim 10$~AU  
(during which the radio core should remain quite steady); further evidence of persistent radio jets beyond the core at
scales $\sim 10-1000$~AU; hints of a spiral-like radio structure (B08) in the still marginally resolved radio core;
some kind of correlation between the jet-like radio emission at distances $<100$~AU and the TeV radiation (which varies
periodically along the orbit; Aharonian et al. \cite{aharonian06}).

\begin{acknowledgements} 
We thank Josep Mart\'i for useful suggestions. 
We thank also an anonymous referee for constructive comments.
V.B-R. gratefully acknowledges support from the Alexander von Humboldt Foundation.
V.B-R. acknowledges support by DGI of MEC under grant
AYA2007-68034-C03-01, as well as partial support by the European Regional Development Fund (ERDF/FEDER).
\end{acknowledgements}

{}


\begin{thebibliography}{}

\bibitem[2005]{aharonian05}
Aharonian, F., Akhperjanian, A.~G., Aye, K.~M., et~al. 2005, Science, 309, 746

\bibitem[2006]{aharonian06}
Aharonian, F., Akhperjanian, A.~G., Bazer-Bachi, A.~R., et~al. 2006, A\&A, 460, 743

\bibitem[2003]{benaglia03}
Benaglia, P. \& Romero, G.~E. 2003, A\&A, 399, 1121

\bibitem[2008]{bogovalov08} 
Bogovalov, S.~V., Khangulyan, D., Koldoba, A.~V., Ustyugova, G.~V., Aharonian, F. 2008, MNRAS, 387, 63

\bibitem[2007]{bosch07}
Bosch-Ramon, V., Motch, C., \& Rib\'o, M., et~al. 2007, A\&A,  473, 545 

\bibitem[2008a]{bosch08}
Bosch-Ramon, V., Khangulyan D., \& Aharonian, F. 2008a, A\&A, 482, 397 (B08)

\bibitem[2008b]{bosch08b}
Bosch-Ramon, V., Khangulyan D., \& Aharonian, F. 2008b, A\&A, 489, L21 

\bibitem[2005]{casares05} 
Casares, J., Rib\'o, M., \& Ribas, I., et al. 2005, MNRAS, 364, 899  (C05)

\bibitem[2006]{dhawan06} 
Dhawan, V., Mioduszewski, \& A., Rupen, M. 2006, 
The VI Microquasar Workshop: Microquasars and Beyond (Proceedings of Science), 52, 1

\bibitem[2006]{dubus06}
Dubus, G. 2006, A\&A,  456, 801

\bibitem[2001]{fender01} 
Fender, R. 2001, MNRAS, 322, 31

\bibitem[2008]{godambe08} 
Godambe, S., Bhattacharya, S., Bhatt, N. \& Manojendu, C. 2008, MNRAS, 390, 43 (G08)

\bibitem[2008]{khangulyan08}
Khangulyan, D., Aharonian, F., \& Bosch-Ramon, V. 2008, MNRAS, 383, 467 

\bibitem[2001]{krtidka01} 
Krtidka, J. \& Kub\'at, J. 2001, A\&A, 377, 175

\bibitem[1998]{marti98}
Mart{\'{\i}}, J., Paredes, J.~M., \& Rib\'o, M. 1998, A\&A, 338, L71 (M98)

\bibitem[2005]{martocchia05}
Martocchia, A., Motch, C., \& Negueruela, I. 2005, A\&A, 430, 245 

\bibitem[2006]{owocki06}
Owocki, S. \& Cohen, D. 2006, ApJ, 648, 565

\bibitem[1970]{Pacholczyk70}
Pacholczyk, A.~G., 1970, Radio Astrophysics, Freeman, San Francisco, CA 

\bibitem[2007]{pandey07}
Pandey, M., Rao, A.~P., Ishwara-Chandra, C.~H., Durouchoux, P., \& Manchanda, R.~K. 2007, A\&A, 463, 567 (P07)

\bibitem[2000]{paredes00}
Paredes, J.M., Mart\'{\i}, J., Rib\'{o}, M.,  Massi, M. 2000, Science, 288, 2340

\bibitem[2002]{paredes02}
Paredes, J.~M., Rib\'o, M., Ros, E., Mart\'i, J., \& Massi, M. 2002, A\&A, 393, L99

\bibitem[2006]{paredes06}
Paredes, J.~M., Bosch-Ramon, V., \& Romero, G.~E. 2006, A\&A, 451, 259 

\bibitem[2008]{perucho08}
Perucho, M. \& Bosch-Ramon, V. 2008, A\&A, 482, 917  

\bibitem[2008]{ribo08}
Rib\'o, M., Paredes, J.~M., Mold\'on, J., Mart\'i, J., \& Massi, M. 2008, A\&A, 481, 17 (R08)

\bibitem[2007]{romero07}
Romero, G.~E., Okazaki, A.~T., Orellana, M., Owocki, S.~P. 2007, A\&A, 474, 15

\bibitem[1979]{rybicki79}
Rybicki, G.~B. \& Lightman, A.~P. 1979, Radiative processes in astrophysics (New York: Wiley-Interscience)

\end{thebibliography}
\end{document}